\documentclass[aps,prd,floatfix,nofootinbib,showpacs,twocolumn,10pt]{revtex4}

\usepackage{amssymb}
\usepackage[intlimits]{amsmath}
\usepackage{amsfonts}
\usepackage{dsfont}
\usepackage{subfigure}
\usepackage{nicefrac}
\usepackage[usenames,dvipsnames]{color}

\definecolor{webgreen}{rgb}{0,0.75,0}
\definecolor{webred}{rgb}{0.75,0,0}
\definecolor{webblue}{rgb}{0,0,0.75}
\definecolor{darkblue}{rgb}{0,0,0.6}
\definecolor{dunkelgrau}{rgb}{0.8,0.8,0.8}
\definecolor{lgray}{rgb}{0.95,0.95,0.95}
\definecolor{lgreen}{rgb}{0.95,1.00,0.90}
\definecolor{lblue}{rgb}{0.9,0.95,1.00}
\definecolor{lred}{rgb}{1.00,0.90,0.80}
\definecolor{shadecolor}{rgb}{1.00,0.92,0.82}

\usepackage{nicefrac}
\usepackage{slashed}

\usepackage[colorlinks=true,linkcolor=darkblue,citecolor=darkblue,
urlcolor=darkblue]{hyperref}
\usepackage{graphicx}

\usepackage{natbib}
\usepackage{multirow}
\usepackage{bbm}

\newcommand{\Slash}[1]{\ooalign{\hfil/\hfil\crcr$#1$}}

\begin{document}

\title{Octet and Decuplet masses: a covariant three-body Faddeev 
calculation}
\author{H\`{e}lios \surname{Sanchis-Alepuz}}
\email{helios.sanchis-alepuz@theo.physik.uni-giessen.de}
\affiliation{Institute of Theoretical Physics, Justus-Liebig University of 
Gie\ss en, Heinrich-Buff-Ring 16, 35392, Gie\ss en, Germany}
\author{Christian S. \surname{Fischer}}
\affiliation{Institute of Theoretical Physics, Justus-Liebig University of 
Gie\ss en, Heinrich-Buff-Ring 16, 35392, Gie\ss en, Germany}

\begin{abstract}
We determine the baryon octet and decuplet masses as well as their 
wave functions in a covariant three-body Faddeev approach. We work 
in an approximation where irreducible three-body forces are neglected.
In the two-body interactions we take into account a well explored
rainbow-ladder kernel as well as flavor dependent meson-exchange terms 
motivated from the underlying quark-gluon interaction. Without flavor
dependent forces we find agreement with the experimental spectrum 
on the 5-10 percent level. Including the flavor dependent terms on 
an exploratory level delivers a $\Sigma-\Lambda$ splitting with the 
correct sign although the magnitude is still too small.
\end{abstract}


\maketitle
\date{\today}


\section{Introduction}\label{sec:introduction}
 
Most of the mass of ordinary matter in the observable Universe is generated by 
the strong interaction. Therefore, it is important to understand the mechanisms 
of mass generation in QCD, the theory of strong interactions. Of particular 
theoretical interest is the calculation of the hadron spectrum from QCD and its 
comparison with the experimentally measured masses. For this reason, great 
effort is put to overcome the technical difficulties of lattice QCD and provide 
reliable \textit{ab initio} calculations of hadron properties (see, e.g. 
\cite{Lin:2008pr,Dudek:2011tt,Edwards:2012fx,Fodor:2012gf,Engel:2011aa,
Engel:2013ig} and references therein).

The combination of Dyson-Schwinger equations (DSE) and Bethe-Salpeter 
equations (BSE) provides, ideally, an 
alternative approach to first principle QCD. Complementary to the 
lattice, it provides insight into the underlying mechanism of mass 
generation in QCD and the details of the interaction mechanisms at
work in binding quarks and gluon together into hadrons. In practice, 
of course, challenges arise. Most prominently for most applications 
the infinite set of DSEs that define a theory must be truncated 
into something manageable, which in turn induces the necessity of modeling. It 
is nevertheless remarkable that, even with the simplest truncations, the 
framework is able to reproduce fairly well many hadron observables (see, e.g. 
\cite{Bashir:2012fs,Eichmann:2013afa} for overviews). Despite this fact, the 
deficiencies of these simple truncations are apparent when the details of the 
interaction are probed, such as in the calculation of baryon form factors 
\cite{Eichmann:2011vu,Nicmorus:2010sd,Sanchis-Alepuz:2013iia} or in meson and 
baryon excited states \cite{Blank:2011ha,Fischer:2014xha,Sanchis-Alepuz:2014mca,Popovici:2014pha,Gomez-Rocha:2014vsa}. 
A major current focus is therefore to improve the approximation of the
quark-gluon interaction in the DSE/BSE system
\cite{Bender:1996bb,Watson:2004kd,Bhagwat:2004hn,Matevosyan:2006bk,Fischer:2009jm,Chang:2009zb,Heupel:2014ina,Fischer:2007ze,Sanchis-Alepuz:2014wea}.

The calculation of bound states, especially of baryons, in the DSE/BSE formalism 
is also technically challenging. A fully covariant three-body 
calculation of the nucleon \cite{Eichmann:2009qa}, delta and omega 
\cite{SanchisAlepuz:2011jn} masses has already been achieved. However, for
baryon octet and decuplet states with both $u/d$- and $s$-quark content 
only a simplified two-body framework using the quark-diquark approximation has 
been considered so far \cite{Oettel:1998bk}. In this work, we overcome those technical 
difficulties and unify, for the first time, the treatment of light and strange 
baryons in the three-body approach. We explore the merits of the simplest truncation 
of the quark-gluon interaction compatible with Poincar\'e covariance and chiral 
symmetry, namely, the rainbow-ladder truncation and provide first steps to
include flavor dependent interactions responsible for details such as the 
$\Sigma-\Lambda$ splitting in the spectrum. The methodology presented here sets 
the stage for future, more sophisticated, schemes beyond the simple rainbow-ladder framework.

The paper is organized as follows. In Section \ref{sec:formalism} we briefly summarize 
the covariant DSE/BSE formalism. A more detailed description of the formalism is given 
in Appendix \ref{sec:review_faddeev}. In Section \ref{sec:results} we discuss the masses 
of the baryon octet and decuplet in the rainbow-ladder truncation and explore the possible 
effect of flavor dependent interactions beyond rainbow-ladder. Potential future developments 
are discussed in the conclusions, Section \ref{sec:summary}. 


\section{Review of the formalism}\label{sec:formalism} 

\begin{figure*}[hbtp]
 \begin{center}
  \includegraphics[width=0.86\textwidth,clip]{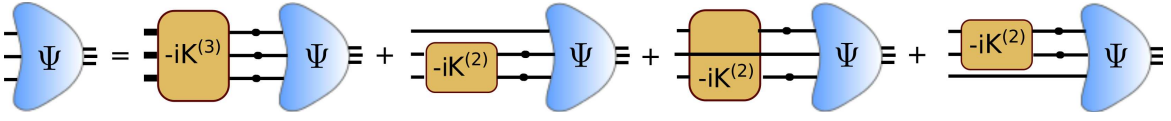}
 \end{center}
 \caption{Diagrammatic representation of the three-body Bethe-Salpeter 
equation with amplitude $\Psi$. Full quark propagators are denoted by straight lines with black dots.
The interaction between the quarks are encoded in the three-body and two-body
kernels $K^{(3)}$ and $K^{(2)}$.}\label{fig:faddeev_eq}
\end{figure*}

In the Bethe-Salpeter framework, a baryon is described by the three-body 
Bethe-Salpeter 
amplitude $\Gamma_{ABCD}(p_1,p_2,p_3)$, where we use $\{ABC\}$ as generic quark 
indices for spin, flavor and color indices (e.g. $A\rightarrow\{\alpha,a,r\}$, 
respectively). The amplitude depends on the three quark momenta $p_{1,2,3}$, 
which can be 
expressed in terms of two relative momenta $p$ and $q$ and the total momentum 
$P$ (see Eq.~\ref{eq:defpq} in Appendix~\ref{sec:review_faddeev}). It is 
decomposed in a tensor product of a spin-momentum part to be 
determined and flavor and color parts which are fixed
  \begin{equation}\label{eq:BSE_amplitude}
 \Gamma_{ABCD}(p,q,P)=\left(\sum_\rho 
\Psi^\rho_{\alpha\beta\gamma\mathcal{I}}(p,q,P) \otimes 
F^\rho_{abcd}\right)\otimes \frac{\epsilon_{rst}}{\sqrt{6}}~.
\end{equation}
The color term $\epsilon_{rst}/\sqrt{6}$ fixes the baryon to be a color singlet 
and the flavor terms $F^\rho_{abcd}$ are the quark-model $SU(3)$-symmetric 
representations (see Appendix~\ref{sec:flavor}). The index $\rho$ 
denotes the 
representation of the $SU(3)$ group to which the baryon belongs 
(mixed-symmetric 
or mixed-antisymmetric representation for the baryon octet and only the 
symmetric representation for the baryon decuplet).

The spin-momentum part of the Bethe-Salpeter amplitude, 
$\Psi^\rho_{\alpha\beta\gamma\mathcal{I}}(p,q,P)$, is a tensor with three Dirac 
indices $\alpha,\beta,\gamma$ associated to the valence quarks and a generic 
index $\mathcal{I}$ whose nature depend on the spin of the resulting bound 
state. They can in turn be expanded in a covariant basis $\{\tau(p,q;P)\}$
\begin{equation}\label{eq:basis_expansion}
\Psi^\rho_{\alpha\beta\gamma\mathcal{I}}(p,q,P)=f^{\rho}_i(p^2,q^2,z_0,z_1,
z_2)\tau_{\alpha\beta\gamma\mathcal{I}}^i(p,q;P)~,
\end{equation}
where the scalar coefficients $\{f\}$ depend on Lorentz scalars $p^2$, $q^2$, 
$z_0=\widehat{p_T}\cdot\widehat{q_T}$, $z_1=\widehat{p}\cdot\widehat{P}$ and 
$z_2=\widehat{q}\cdot\widehat{P}$ only. The subscript $T$ denotes transverse 
projection with respect to the total momentum and vectors with hat are 
normalized. A covariant basis can be obtained using symmetry requirements only; 
for positive-parity spin-$\nicefrac{1}{2}$ baryons it contains 64 elements 
\cite{Eichmann:2009en,Eichmann:2009zx} whereas for spin-$\nicefrac{3}{2}$ 
baryons it contains 128 elements \cite{SanchisAlepuz:2011jn}. The task of 
solving the bound-state equation is now greatly facilitated, as one only needs 
to solve for the scalar functions $f$.

The amplitudes $\Psi$ are, in general, solutions of a three-body BSE (see Fig. 
\ref{fig:faddeev_eq}), for which one needs to specify the three-particle  and 
two-particle irreducible kernels, $K^{(3)}$ and  $K^{(2)}$, respectively. In the 
Faddeev approximation the three-body irreducible interactions are 
neglected, and we refer to the simplified BSE as the 
Faddeev equation (FE). Central elements of the Faddeev 
equation are the full quark propagator $S$ (omitting now Dirac indices)
\begin{equation}\label{eq:full_quark}
 S^{-1}(p)=A(p^2)\left(i\Slash{p}+B(p^2)/A(p^2)\right)~,
\end{equation}
with vector and scalar dressing functions $A(p^2)$ and $B(p^2)$. The ratio 
$M(p^2)=B(p^2)/A(p^2)$ is a renormalisation group invariant and describes the
running of the quark mass with momentum. The dressing functions are obtained as solutions of 
the quark DSE 
\begin{equation}\label{eq:quarkDSE}
 S^{-1}(p)=S^{-1}_{0}(p)+Z_{1f}g^2 C_F \!\!\int_q \!\gamma^\mu
D_{\mu\nu}(p-q)\Gamma^\nu(p,q)S(q)\,\,,
\end{equation}
which also contains the full quark-gluon vertex $\Gamma^\nu$ and the full gluon 
propagator $D_{\mu\nu}$; 
$S_{0}$ is the (renormalized) bare propagator with inverse
\begin{equation}\label{eq:bare_prop}
 S_{0}^{-1}(p)=Z_2\left(i\Slash{p}+m\right)\,,
\end{equation}
where $m$ is the bare quark mass 
and $Z_{1f}$ and $Z_2$ are renormalization constants. The renormalized strong 
coupling is denoted by $g$ and $\int_q = \int \frac{d^4 q}{(2 \pi)^4}$ abbreviates
a four-dimensional integral in momentum space supplemented with a translationally
invariant regularization scheme. 

Under certain symmetry requirements, the practical solution of the Faddeev 
equation 
can be greatly simplified by relating the three two-body interaction diagrams 
in 
Fig.~\ref{fig:faddeev_eq}. 
As shown in Appendix~\ref{sec:review_faddeev}, taking the flavor part of the 
Faddeev amplitudes as representations of the $SU(3)$ group induces a specific 
transformation rule for the spin-momentum part under interchange of its 
valence-quark indices. In this case the Faddeev equation for the coefficients 
$f$ reduces to
\begin{flalign}\label{eq:Faddeev_coeff}
 f^{\rho}_i(p^2,q^2,z_0,z_1,z_2)=& \nonumber\\        
C\mathcal{F}_1^{\rho\rho';\lambda}~H_1^{ij}&~g^{\rho'',\lambda}_{j}(p'^2,q'^2,
z'_0,z'_1,z'_2)+ \nonumber\\
C\mathcal{F}_2^{\rho\rho';\lambda}~H_2^{ij}&~g^{\rho'',\lambda}_{j}(p''^2,q''^2,
z''_0,z''_1,z''_2)+ \nonumber\\
C\mathcal{F}_3^{\rho\rho';\lambda}~&~g^{\rho',\lambda}_{i}(p^2,q^2,z_0,z_1,z_2)~,
\end{flalign}
with the color factor $C$, the flavor matrices $\mathcal{F}$, the rotation 
matrices $H$ and other symbols defined in Appendix \ref{sec:review_faddeev}. 
Here, the index $\lambda$ runs over all elements in a given flavor state (e.g. 
$\lambda=1,2$ for the mixed-symmetric flavor wave function of the proton and 
$\lambda=1,3$ for the mixed-antisymmetric one; see Appendix \ref{sec:flavor}). 
The simplification is that one needs to solve for one of the diagrams only
\begin{widetext}
\begin{flalign}\label{eq:Faddeev_coeff3}
g^{\rho,\lambda}_i(p^2,q^2,z_0,z_1,z_2)=\int_k~\textnormal{Tr}\Bigl[~\bar{\tau}^
{i}_{\beta\alpha\mathcal{I}\gamma}(p,q,P)K_{\alpha\alpha',\beta\beta'}(p,q,
k)~&\delta_{\gamma\gamma''}S^{(\lambda_1)}_{\alpha'\alpha''}(k_1)~S^{(\lambda_2)
}_{\beta'\beta''}(k_2)~\tau^{j}_{\alpha''\beta''\gamma''\mathcal{I}}(p_{(3)},q_{
(3)},P)~\Bigr]\times \nonumber\\& 
f^{\rho}_j(p_{(3)}^2,q_{(3)}^2,z^{(3)}_0,z^{(3)}_1,z^{(3)}_2)~.
\end{flalign}
\end{widetext}
where now the quark at the position $\ell$ in each term of the flavor wave 
function is denoted by the superindex $\lambda_\ell$. The internal relative 
momenta $p_{(3)}$, $q_{(3)}$ (and analogously for $z^{(3)}_0$, $z^{(3)}_1$ and 
$z^{(3)}_2$)  are obtained from (\ref{eq:defpq}) upon substitution of the quark 
momenta by the internal counterparts $k_1=p_1-k$ and $k_2=p_2+k$. The conjugate 
of the covariant basis $\bar{\tau}$ has been defined in 
\cite{Eichmann:2009en,SanchisAlepuz:2011jn} and it is assumed that the basis 
$\{\tau\}$ is orthonormal.

So far we have not specified the two-body kernel $K_{\alpha\alpha',\beta\beta'}$ 
and, in fact, the simplification (\ref{eq:Faddeev_coeff}), 
(\ref{eq:Faddeev_coeff3}) applies for any kernel provided that 
$K_{\alpha\alpha',\beta\beta'}=K_{\beta\beta',\alpha\alpha'}$. However, the 
kernel contains all possible two-body irreducible interactions among two quarks 
and for any practical implementation of Eq. (\ref{eq:Faddeev_coeff3}) one must 
truncate it. The two-body kernel is related to the integration kernel in the 
quark DSE via the requirements of chiral symmetry expressed by axialvector and
vector Ward-Takahashi identities \cite{Fukuda:1987su,McKay:1989rk,Munczek:1994zz}, 
leading to massless pions in the chiral limit from the pseudoscalar-meson BSE. 
Using crossing symmetry, the quark-antiquark kernel is
then related to the quark-quark kernel that appears in the Faddeev equation.
With these restrictions, if the full gluon propagator and full quark-gluon 
vertex in (\ref{eq:quarkDSE}) are truncated to their tree-level part, the 
corresponding two-body kernel is a single-gluon exchange with a tree-level 
vector-vector coupling to quarks. All non-perturbative effects from both the 
gluon and the vertex are 
encoded in an effective coupling $\alpha_{\textrm{eff}}$ which has to be 
modeled. This is the rainbow-ladder truncation of the DSE/BSE system.
In a series of works \cite{Fischer:2007ze,Sanchis-Alepuz:2014wea}, this 
truncation has been extended to include effects from the four-quark
Green's function in the quark-gluon interaction parameterized in terms 
of (off-shell) meson exchange. In the DSE and Bethe-Salpeter kernel, 
theses effects can be represented by one-pion exchange between the 
quarks while still maintaining the pseudo-Goldstone nature of the pion,
see \cite{Fischer:2007ze} for details. In the following we specify the
rainbow-ladder gluon and pion exchange parts of the resulting interaction 
and then discuss our results.

\subsection{Effective coupling of one-gluon exchange}\label{sec:MT}

For the effective interaction in the rainbow-ladder truncation we use 
the Maris-Tandy model~\cite{Maris:1997tm,Maris:1999nt} which has been employed 
frequently in hadron studies within the rainbow-ladder BSE/DSE framework. 
Despite its simplicity it performs very well when it comes to purely 
phenomenological calculations of ground-state meson and baryon properties
in certain channels, see e.g. \cite{Fischer:2014xha} of a corresponding 
discussion. It combines the relevant parts in the quark self-energy according to
\begin{align}
Z_{1f} C_F\frac{g^2}{4\pi} D_{\mu\nu}(k)\Gamma^\nu(p,q) = Z_2^2C_FT_{\mu\nu}^k 
\frac{\alpha_{\mathrm{eff}}(k^2)}{k^2}\gamma^\nu\;,
\end{align}
such that the resulting kernel in the Baryon Faddeev equation is given by
\begin{equation}\label{eq:RLkernel}
	{K}^{RL}_{\alpha\alpha'\beta'\beta}(k)= -4\pi C_F~Z_2^2
~\frac{\alpha_{\textrm{eff}}(k^2)}{k^2}~
	T_{\mu\nu}(k)~\gamma^\mu_{\alpha\alpha'}  \gamma^\nu_{\beta'\beta}\,\,.
\end{equation}
with gluon momentum $k=p-q$, transverse projector $T_{\mu\nu}^k$ and the 
effective running coupling $\alpha_{\textrm{eff}}$ given by
\begin{flalign}\label{eq:MTmodel}
\alpha_{\textrm{eff}}(k^2) {}=&
 \pi\eta^7\left(\frac{k^2}{\Lambda^2}\right)^2
e^{-\eta^2\frac{k^2}{\Lambda^2}}\nonumber\\ &+{}\frac{2\pi\gamma_m
\big(1-e^{-k^2/\Lambda_{t}^2}\big)}{\textnormal{ln}[e^2-1+(1+k^2/\Lambda_{QCD}
^2)^2]}\,. 
\end{flalign}
This interaction reproduces the one-loop QCD behavior of the quark propagator at 
large momenta 
and features a Gaussian distribution of interaction strength in the intermediate 
momentum region 
that provides for dynamical chiral symmetry breaking. The
scale $\Lambda_t=1$~GeV is introduced for technical reasons and has no impact on
the results. Therefore, the interaction strength is characterized by an energy
scale $\Lambda$ and a dimensionless parameter $\eta$ that controls the width of 
the interaction. 
For the anomalous dimension we use $\gamma_m=12/(11N_C-2N_f)=12/25$,
corresponding to $N_f=4$ flavors and $N_c=3$ colors. For the QCD scale
$\Lambda_{QCD}=0.234$ GeV. 

The scale $\Lambda=0.72$~GeV is adjusted to reproduce the 
experimental pion decay constant from the truncated pion BSE.
This as well as several other pseudoscalar ground-state observables turn out to 
be insensitive to the value of $\eta$ in the range of values of $\eta$ 
between $1.6$ and $2.0$ see, e.g. 
\cite{Krassnigg:2009zh,Nicmorus:2010mc,Eichmann:2011vu}). Moreover, the 
current-quark masses are 
fixed to reproduce the physical pion (for the u/d quarks) and kaon (for the s 
quark) masses. The corresponding values are $m_{u/d}(\mu^2)=3.7$~MeV and 
$m_s(\mu^2)=85$~MeV. The renormalisation scale is chosen to be $\mu^2= (19 
\,\mbox{GeV})^2$.

\begin{table*}[tb!]
 \begin{center}
 \small
\renewcommand{\arraystretch}{1.2}
  \begin{tabular}{|c||c|c|c|c|}\hline
  $\nicefrac{1}{2}^+$     	&  $N$     	& $\Sigma$ 	& $\Lambda$  	
& $\Xi$ \\ \hline\hline
 Faddeev				& 0.930 (3) 	& 1.073 (1) 	& 1.073 
(1) 	& 1.235 (5) \\ \hline
 Experiment  	 		& 0.938			& 1.189			
& 1.116   		& 1.315		\\ \hline
 Relative difference	& $<$ 1 \%		& 10 \%			& 4 \%   
		& 6 \%   	\\ 
\hline
\end{tabular}
\caption{Positive-parity baryon octet masses (in GeV) at the physical point 
from 
the rainbow-ladder truncated Faddeev equation. 
We give the central value of the bands corresponding to a variation of 
$\eta$ between $1.6 \le \eta \le 2.0$ with the halfwidth of the bands added in 
brackets. 
We compare also with experimental values \cite{Beringer:1900zz}.
\label{tab:masses_octet}}
 \end{center}
\end{table*}

\begin{table*}[h!t]
 \begin{center}
 \small
\renewcommand{\arraystretch}{1.2}
  \begin{tabular}{|c||c|c|c|c|}\hline
  $\nicefrac{3}{2}^+$     	&  $\Delta$     & $\Sigma^*$ 	& $\Xi^*$  	
& $\Omega$ \\ \hline\hline
 Faddeev				& 1.21 (2) 	& 1.33 (2) 	& 1.47 
(3) 	& 1.65 (4) \\ \hline
 Experiment 			& 1.232 (1)	& 1.385 (2)	& 1.533 (2) & 
1.672	\\ \hline
 Relative difference 	&  2 \%		&  4 \%		& 4 \%   	& 1 \%   
	\\ \hline
\end{tabular}
\caption{Positive-parity baryon decuplet masses (in GeV) at the physical point 
from the rainbow-ladder truncated Faddeev equation. 
We give the central value of the bands corresponding to a variation of 
$\eta$ between $1.6 \le \eta \le 2.0$ with the halfwidth of the bands added in 
brackets. 
We compare also with experimental values 
\cite{Beringer:1900zz}.\label{tab:masses_decuplet}}
 \end{center}
\end{table*}

\subsection{Flavor dependent part of the interaction}\label{sec:pion}

In Refs.~\cite{Fischer:2007ze,Sanchis-Alepuz:2014wea} the explicit construction,
the chiral properties and some consequences for meson and baryon spectra of pion
contributions to the quark-quark interaction have been explored. Since all 
technical details of this beyond rainbow-ladder framework have been discussed 
at several places already, we refrain from repeating the details here and merely
state the resulting kernel for the exchange of a pion between two quark lines
\begin{align}\label{eq:PiKernel} 
  {K}_{\alpha\alpha'\beta\beta'}^{\textrm{pion}}&(l_1,l_2,l_3,l_4;P)={} \nonumber\\
  &\frac{1}{2}
      [\Gamma^j_{\pi}]_{\alpha\alpha'}\left(\frac{l_1+l_2}{2};P\right)
      [Z_2 \tau^j \gamma_5]_{\beta\beta'}
       D_{\pi}(P) \nonumber\\
 & +\frac{1}{2}
      [Z_2 \tau^j \gamma_5]_{\alpha\alpha'}
      [\Gamma^j_{\pi}]_{\beta\beta'}\left( \frac{l_3+l_4}{2};P\right)
      D_{\pi}(P)    
      \,.
\end{align}
Here $l_{1..4}$ are the incoming and outgoing quark momenta, $\Gamma^j_{\pi}(p,P)$ is the pion Bethe-Salpeter amplitude, with $p$ the relative momentum and $P$ the total momentum, and $D_{\pi}$ is the on-shell pion propagator
\begin{equation}\label{eq:pion_propagator}
 D_{\pi}(P)=\frac{1}{M_{\pi}^2+P^2}~.
\end{equation}
Including the pion exchange kernel in the Faddeev equation the resulting total 
kernel is then given by the sum of the rainbow ladder part and the pion exchange part,
\begin{align}
{K}_{\alpha\alpha'\beta\beta'}&={K}_{\alpha\alpha'\beta\beta'}^{\textrm{RL}}+{K}_{\alpha\alpha'\beta\beta'}^{\textrm{pion}}\,.
\end{align}
The same is true for he quark-DSE; see Refs.~\cite{Fischer:2007ze,Sanchis-Alepuz:2014wea} for details.

We wish to emphasize some general points related to this interaction. First, 
we stress that the pion in Eq.~(\ref{eq:PiKernel}) is not an elementary field. 
The pion propagator together with the Bethe-Salpeter amplitudes  
\begin{eqnarray}
\Gamma_\pi^j(p;P) &=& \tau^j \gamma_{5} \left[E_\pi(p;P)- i \slashed P F_\pi(p;P) \right.\nonumber\\
&&\left.
- i \slashed p G_\pi(p;P) - \left[\slashed P ,\slashed p \right] H_\pi(p;P)\right]\,.\label{eq:pseu}
\end{eqnarray}
describe the propagation of a quark anti-quark bound state and its
coupling to two quark lines. The Bethe-Salpeter amplitudes are determined 
by the pion-BSE with the corresponding interaction kernel, i.e. one-gluon 
and pion exchange. Thus the coupling strength of the pion to the quark is 
not a separate model input, but in principle follows from a self-consistent
calculation. In practice, one may resort to suitable approximations as 
detailed below. In general, however, the combined interaction of one-gluon 
plus pion exchange is derived and motivated from the structure of the QCD
Dyson-Schwinger equations \cite{Fischer:2007ze} and not some extra element
added 'on top of QCD'.


\section{Results}\label{sec:results}

\subsection{Spectrum for one-gluon exchange}\label{subsec:masses}

We start our investigation using the one-gluon exchange part of the 
rainbow-ladder kernel, Eq.~(\ref{eq:RLkernel}), only. This part is 
flavor blind, and as a consequence  
the bound state masses are determined by their quark content only. 
That is, although the symmetry 
properties of the flavor part of the Faddeev amplitude are important to relate 
all three diagrams in the Faddeev equation, this flavor part plays no further 
role in the calculation of the spectrum in the present truncation. As a 
consequence, in the baryon octet, the $\Lambda$ and $\Sigma$ baryons 
are degenerate. Moreover, since we use the same current-quark mass for the 
$u/d$-quarks, all members in the same isospin multiplet are degenerate.

\begin{table*}[th!]
 \begin{center}
 \small
\renewcommand{\arraystretch}{1.2}
  \begin{tabular}{|c|c|}\hline
 $\Sigma$/$\Lambda$ 	& $\Xi$  \\ \hline\hline
  1.07 (1) 	& 1.22 (1)  \\ 
\hline
\end{tabular}
\hspace*{1cm}
\begin{tabular}{|c|c|c|}\hline
 $\Sigma^*$ 	& $\Xi^*$  	& $\Omega$ \\ \hline\hline
  1.32 (2) 	& 1.47 (2) 	& 1.64 (4)  \\ 
\hline
\end{tabular}
\caption{Positive-parity strange-baryon octet (left) and decuplet (right) masses 
(in GeV) at the physical point 
from the modified (quark-mass dependent) effective interaction.
We give the central value of the bands corresponding to a variation of 
$\eta$ between $1.6 \le \eta \le 2.0$ with the halfwidth of the bands added in 
brackets.\label{tab:masses_mixed}}
 \end{center}
\end{table*}

Table \ref{tab:masses_octet} shows the calculated masses of the three 
representatives of the 
baryon octet and compares them to the experimental values. As an estimate of the 
model 
uncertainty, we calculate for a fixed value of the parameter $\Lambda$ and for a 
range of values from $\eta=1.6$ to $\eta=2.0$ 
as discussed above; we give the central value of this band together with the 
range of results 
indicated in brackets. 
The result for the nucleon is not new and was presented for the first time in 
\cite{Eichmann:2009qa}. It is in excellent agreement with the experimental 
value. To fully appreciate that, one must recall that not only the truncation 
and model used are among the simplest one can define, but also that the model 
parameters were fit in the pseudoscalar meson sector. Still, the agreement in 
the baryon sector is remarkable. For baryons with one strange quark 
the mass obtained is lower than the experimental value for both 
the $\Sigma$ multiplet and the $\Lambda$ singlet. However, the error is reasonably 
small, namely $\sim$10\% if compared with the $\Sigma$ and $\sim$4\% if 
compared with the $\Lambda$. The same qualitative behavior appears for the 
$\Xi$ baryons (or $uss$ state), where we find a $\sim$6\% deviation as
compared to experiment.

In table \ref{tab:masses_decuplet} we show the results of the corresponding 
representatives in the baryon decuplet. The states with equal quark content, 
namely the $\Delta$ and the $\Omega$ were already given in 
\cite{SanchisAlepuz:2011jn}\footnote{The difference between the results 
presented here, especially for the $\Delta$ and the $\Omega$, and those shown in 
previous calculations are due to an improvement in the numerics. In particular 
we have managed to use more angular points in our numerical integration.}. The 
agreement with experiment in these cases is also remarkable. For the newly 
calculated spin-$\nicefrac{3}{2}$ states, $\Sigma^*$ ($uus$) and $\Xi^*$ ($uss$), we 
observe the same qualitative behavior as for their octet counterparts. The 
resulting masses are below the experimental values, with an error of only 
$\sim$4\% in both cases.

Despite these relatively small differences between our results and the 
experimental values a clear deficiency of the rainbow-ladder truncation 
is manifest: there is no mechanism to generate a mass splitting between 
the $\Lambda$ and $\Sigma$ baryons. In order to improve this situation
a flavor dependent interaction is necessary for a more complete description 
of the octet and decuplet baryons. 
In this respect one can think of two different categories of flavor-dependent 
two-body kernels. The first one would have a quark-mass dependence manifest
already in the effective coupling of the rainbow-ladder kernel reflecting the
underlying mass dependence of the quark line going through the quark-gluon 
vertex. 
We test 
this heuristically in the next section through a minimal modification of the 
effective interaction. The second possibility would be a kernel with an explicit 
non-diagonal flavor part which, in particular, would distinguish between the 
$\Lambda$ and $\Sigma$ baryons. An exploratory study in this direction is 
performed in Section \ref{subsec:pion_exchange}.

%

\subsection{Exploration of a quark-mass dependent 
interaction}\label{subsec:mass_dependent_MT}

\begin{figure}[b]
 \begin{center}
  \includegraphics[width=0.45\textwidth]{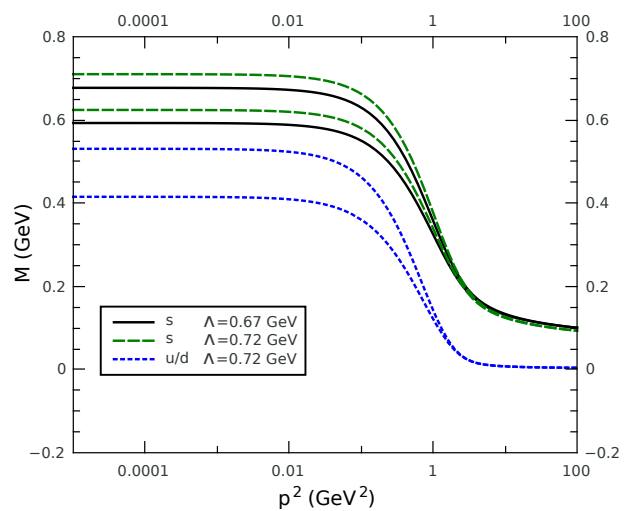}
 \end{center}
 \caption{(color online). Quark mass function as function of the squared 
momentum.}\label{fig:constituent_mass_mixedMT}
\end{figure}

In this section we investigate, in a very simplistic manner, the possible 
effects of a mass dependence in the model used as the effective interaction in a 
rainbow-ladder truncation. In a recent calculation \cite{Williams:2014iea} the 
full quark-gluon vertex has been solved using its DSE under certain truncations. 
There it is observed that the dressing accompanying the tree-level component 
shows a significant quark-mass dependence, becoming weaker as the quark mass 
increases. 

In order to mimic this behavior, we allow the strength parameter $\Lambda$ of 
the 
one-gluon exchange part of the interaction, Eq.~(\ref{eq:MTmodel}), to depend on 
the 
quark flavor. 
%
%
We proceed as follows: 
for $u/d$ quarks we use the same value $\Lambda=0.72$~GeV as in the 
previous section; for the $s$ quark we weaken the interaction taking 
$\Lambda=0.67$~GeV instead in line with the results of \cite{Williams:2014iea}. 
In order to reproduce correctly the kaon mass we 
change the $s$-quark current mass to $m_s(19~\mbox{GeV})=91.5$~MeV. The 
resulting values for the kaon mass and kaon decay constant are $496$~MeV and 
$106$~MeV, respectively. Also, with this choice of parameters, the mass of a 
fictitious $s\overline{s}$ pseudoscalar meson increases by about $\sim 30$~MeV. In the 
present case we also allow the parameter $\eta$ to vary between $1.6$ and $2.0$ 
to estimate the model dependence.

The resulting strange-baryon masses for both the octet and decuplet are 
presented 
in Table \ref{tab:masses_mixed}. Contrary to what might be expected, the masses 
of 
all states remain virtually unchanged as compared to the results presented in 
section \ref{subsec:masses}
above. The reason for this is a compensation mechanism: 
since the same interaction model is used in both 
the Faddeev equation and the quark DSE, a weaker binding in the former is 
accompanied by less dynamical quark generation from the latter. For the quark
this can be seen 
in Fig.~\ref{fig:constituent_mass_mixedMT} where we plot the mass function 
for both the original and the modified effective interactions. The decrease
in quark mass together with the weaker binding in the Faddeev kernel compensate
each other.

Within a rainbow-ladder truncation, a quark-mass dependent effective interaction 
is the only \textit{new} feature that one could incorporate in the 
framework. We have seen that, for baryons, this doesn't improve the comparison 
with experiment. This is an indication that important corrections must rather 
come from missing structures in our interaction kernel. This is explored
in the next section.
%

\subsection{Exploration of a flavor non-diagonal 
interaction}\label{subsec:pion_exchange}

In the rainbow-ladder truncation used so far, as well as in any other truncation with a flavor-independent kernel, the flavor matrices $\mathcal{F}$ appearing in (\ref{eq:Faddeev_coeff}) and defined in (\ref{eq:flavor_matrices}) are the same for any state in the same $SU(3)$ multiplet. In particular, this means that the spin-$\nicefrac{1}{2}$ $\Lambda$ and $\Sigma$ baryons, with equal quark content but different symmetry properties of their flavor amplitudes, will have the same mass and other properties. If, instead, the interaction kernel features a flavor dependence via the mixing of quark legs, the flavor matrices will be different for different states. 
Such an interaction kernel has been specified in section \ref{sec:pion}. Note that the pion couples
to the quark lines according to its flavor content encoded in the Pauli matrices $\tau$. 
These generate the flavor mixing. In particular, this kernel generates the following flavor matrices for the lambda
\begin{flalign}
\mathcal{F}_1&=\left(\begin{array}{cc}
                     \frac{1}{2} & 0 \\
                     \frac{-\sqrt{3}}{2} & 0 
                    \end{array}\right)~, \hspace*{1cm}
\mathcal{F}_2=\left(\begin{array}{cc}
                     \frac{1}{2} & 0 \\
                     \frac{\sqrt{3}}{2} & 0 
                    \end{array}\right)~, \\
\mathcal{F}_3&=\left(\begin{array}{cc}
                     -1 & 0 \\
                      0 & 0 
                    \end{array}\right)~,\nonumber
\end{flalign}
and for the sigma
\begin{flalign}
\mathcal{F}_1&=\left(\begin{array}{cc}
                     0 & \frac{-1}{2\sqrt{3}} \\
                     0 & \frac{-1}{6} 
                    \end{array}\right)~, \hspace*{1cm}
\mathcal{F}_2=\left(\begin{array}{cc}
                     0 & \frac{1}{2\sqrt{3}} \\
                     0 & \frac{-1}{6} 
                    \end{array}\right)~,  \\
\mathcal{F}_3&=\left(\begin{array}{cc}
                      0 & 0 \\
                      0 & \frac{1}{3} 
                    \end{array}\right)~.\nonumber
\end{flalign}

A self-consistent calculation, where the quark DSE, the pion BSE and the Faddeev equation are all solved with the inclusion of the kernel (\ref{eq:PiKernel}) was presented in \cite{Sanchis-Alepuz:2014wea} for the case of the nucleon and $\Delta$ baryons. In the case of the baryon octet and decuplet, a self-consistent calculation would imply solving the DSE/BSE system with an extension of the kernel (\ref{eq:PiKernel}) in which the exchange of any member of the pseudoscalar meson octet is allowed. This is beyond the scope of the present work. 
However, we can illustrate the effect of these type of interactions with the
pion exchange only, i.e. Eq.~(\ref{eq:PiKernel}), in a minimal approximation
where we replace the pion amplitude $\Gamma_{\pi}$, by the exact solution of
its leading amplitude in the chiral limit $\Gamma_{\pi}(p,P)=\gamma_5 B(p^2)/f_{\pi}$.
Here $f_{\pi}$ is the pion decay constant and $B(p^2)$ the quark dressing 
function (\ref{eq:full_quark}). 
Following \cite{Sanchis-Alepuz:2014wea} we use a higher interaction parameter $\Lambda=0.84$, but we maintain here the aforementioned quark masses. The resulting $\Lambda$ and $\Sigma$ masses are shown in Table \ref{tab:masses_pion_exchange}. 
\begin{table}[bt!]
 \begin{center}
 \small
\renewcommand{\arraystretch}{1.2}
  \begin{tabular}{|c|c|}\hline
 $\Lambda$ & $\Sigma$  \\ \hline\hline
  1.161 (7) 	& 1.164 (9)  \\   
\hline
\end{tabular}
\caption{Lambda and Sigma baryon masses 
(in GeV) at the physical point 
from the flavor dependent kernel (\ref{eq:PiKernel}), using $B/f_{\pi}$ for the full pion amplitude. We give the central value of the bands corresponding to a variation of 
$\eta$ between $1.6 \le \eta \le 2.0$ with the halfwidth of the bands added in 
brackets.\label{tab:masses_pion_exchange}}
 \end{center}
\end{table}
Although this simplified interaction generates an exceedingly small splitting\footnote{To ensure that this is not a purely numerical effect, we have run the calculation enhancing the effect of (\ref{eq:PiKernel}) by an artificial large factor. We confirmed that the splitting is indeed an effect of the flavor-mixing kernel.}, 
it has the correct sign and therefore serves to illustrate what kind of interaction kernels must be included in order to reproduce correctly the baryon octet spectrum. 

As such, this not a new result. Flavor dependent interaction terms have been 
discussed in great detail in the context of quark model calculations; see e.g.
the reviews \cite{Glozman:1995fu,Capstick:2000qj,Klempt:2009pi} and 
Refs. therein. The novelty of our approach lies in the fact that our framework
is entirely based on QCD. As already discussed above, the flavor dependent 
interaction terms are not introduced by adding additional fundamental meson 
fields to the theory but appear self-consistently as a result of dynamical 
chiral symmetry breaking and the formation of colorless bound states that 
in turn may contribute to QCD correlation functions. We believe this is an
important conceptual progress as compared to previous approaches.


\section{Summary}\label{sec:summary}

We presented a fully covariant three-body calculation of the octet 
and decuplet baryon spectrum in the rainbow-ladder truncated DSE/BSE 
approach. We showed that a simple rainbow-ladder like interaction is
sufficient to reproduce the mass spectrum of the octet and the decuplet
on a level better than ten percent. However, when it comes to the fine
details, the deficiencies of this type of interaction become apparent.
In particular, it was necessary to include flavour non-diagonal 
interaction terms to induce a splitting between the $\Lambda$ and $\Sigma$
baryons of the octet and decuplet. These terms have been motivated and derived
from the underlying quark-gluon interaction in previous work 
\cite{Fischer:2007ze,Sanchis-Alepuz:2014wea}. In principle, although with 
much increased numerical effort, it is possible to include these interaction 
terms in a much more complete manner. This will be the subject of future work.


\acknowledgments
We thank Richard Williams, Walter Heupel and Gernot Eichmann for helpful discussions.

This work has been supported by an Erwin Schr\"dinger fellowship J3392-N20 
from the Austrian Science Fund, FWF, by the Helmholtz International Center 
for FAIR within the LOEWE program of the State of Hesse, and by the 
DFG collaborative research center TR 16.

\appendix


\section{Faddeev equation with different flavors}\label{sec:review_faddeev}

The method described in this appendix was first introduced in 
\cite{Eichmann:2011vu} for the nucleon Faddeev equation and applied in 
\cite{SanchisAlepuz:2011jn} for the delta and omega. Here, we review its main 
aspects making explicit the possibility of having quarks with different masses 
in the problem.

Being a representation of a three-quark Green's function, the three-body 
Bethe-Salpeter amplitude (\ref{eq:BSE_amplitude}) must be antisymmetric in the 
quark indices $\{A,B,C\}$. The color part $\epsilon_{rst}$ is already 
antisymmetric in order to restric the state to be a color singlet, which implies 
that the product of the flavor and spin-momentum parts must be symmetric. 
Assuming flavor-$SU(3)$ symmetry (the same symmetry arguments would hold, of 
course, if we take the flavor symmetry group larger) one constructs the usual 
octet and decuplet representations for baryons (see, e.g., \cite{Leader:1996hk} 
and Appendix \ref{sec:flavor}). There are two flavor octets, one symmetric and 
one antisymmetric with respect to the first two quark indices. The physical 
octet is a quantum superposition of these two. There is only one flavor 
decuplet, whose states are symmetric in all their indices. That is, each baryon 
is associated to a flavor wave-function $F^\rho_{abc}$, with $\rho=1,2$ for the 
antisymmetric and symmetric octet representations, 
respectively, and $\rho=1$ for the decuplet representations. We will be 
interested here in the quark permutations $(123)\rightarrow (231)$ and 
$(123)\rightarrow (312)$, under which the product of spin-momentum and flavor 
parts transforms symmetrically
\begin{align}\label{eq:spin_flavor_transformation} 
\Psi^\rho_{\alpha\beta\gamma\mathcal{I}}(p_1,p_2,p_3)~F^\rho_{abcd}&=\Psi^\rho_{
\beta\gamma\alpha\mathcal{I}}(p_2,p_3,p_1)~F^{\rho'}_{bcad}~, \nonumber\\
\Psi^\rho_{\alpha\beta\gamma\mathcal{I}}(p_1,p_2,p_3)~F^\rho_{abcd}&=\Psi^\rho_{
\gamma\alpha\beta\mathcal{I}}(p_3,p_1,p_2)~F^{\rho'}_{cabd}~.
\end{align}

In general, the Faddeev equation reads
\begin{widetext}

  \begin{flalign}\label{eq:symbolic_faddeev_original}
   \Gamma_{ABC 
\mathcal{I}}(p,q,P)={}&\int_k~\Bigl[~\textrm{K}_{BB',CC'}(k_2,\widetilde{k}
_3;k)~\delta_
{AA''}~\textrm{S}_{B'B''}(k_2)~\textrm{S}_{C'C''}(\widetilde{k}_3)\Gamma_{
A''B''C''\mathcal{I}}(p^{(1)},q^{(1)},P)~\Bigr]~+ \nonumber \\                      
&\int_k~\Bigl[~\textrm{K}_{AA',CC'}(k_3,\widetilde{k}_1;k)~\delta_
{BB''}~\textrm{S}_{A'A''}(\widetilde{k}_1)~\textrm{S}_{C'C''}(k_3)\Gamma_{
A''B''C''\mathcal{I}}(p^{(2)},q^{(2)},P)~\Bigr]~+\nonumber\\
&\int_k~\Bigl[~\textrm{K}_{AA',BB'}(k_1,\widetilde{k}_2;k)~\delta_
{CC''}~\textrm{S}_{A'A''}(k_1)~\textrm{S}_{B'B''}(\widetilde{k}_2)\Gamma_{
A''B''C''\mathcal{I}}(p^{(3)},q^{(3)},P)~\Bigr]~,                   
  \end{flalign}
  \end{widetext}
with
\begin{align}\label{eq:defpq}
        p &= (1-\zeta)\,p_3 - \zeta (p_1+p_2)\,, &  p_1 &=  -q -\dfrac{p}{2} +
\dfrac{1-\zeta}{2} P\,, \nonumber\\
        q &= \dfrac{p_2-p_1}{2}\,,         &  p_2 &=   q -\dfrac{p}{2} +
\dfrac{1-\zeta}{2} P\,, \nonumber\\
        P &= p_1+p_2+p_3\,,                &  p_3 &=   p + \zeta  
P\,\,,\nonumber\\
\end{align}
and $\zeta=1/3$ a momentum partitioning parameter. The internal quark 
propagators depend on 
the internal quark momenta
$k_i=p_i-k$ and $\tilde{k}_i=p_i+k$, with $k$ the exchanged momentum. The 
internal
relative momenta, for each of the three terms in the Faddeev equation, are
\begin{align}\label{internal-relative-momenta}
p^{(1)} &= p+k,& p^{(2)} &= p-k,& p^{(3)} &= p,\nonumber\\
q^{(1)} &= q-k/2,& q^{(2)} &= q-k/2, & q^{(3)} &= q+k\,\,.\nonumber\\
\end{align}
The kernel contains spin-momentum, flavor and color parts as well
\begin{equation} 
\textrm{K}_{AA',BB'}(p_1,p_2;p)=K_{\alpha\alpha',\beta\beta'}(p_1,p_2;p)k^F_{
aa'bb'}k^C_{rr',ss'}~.
\end{equation}
Moreover, we write the flavor wave functions as a sum of several terms
\begin{equation}
 F^\rho_{abcd}=\sum_\lambda^{d_F} F^{\rho,\lambda}_{abcd}
\end{equation}
where $d_F$ is the number of such terms (for example, the antisymmetric 
component of the $\Xi^0$ is $(uss-sus)/\sqrt{2}$ and therefore $d_F=2$). The 
action of the propagators on the Faddeev amplitudes in 
(\ref{eq:symbolic_faddeev_original}) can therefore be written as a sum of $d_F$ 
terms, for instance
\begin{flalign} 
\textrm{S}_{AA'}(p_1)&\textrm{S}_{BB'}(p_2)\Gamma_{A'B'C\mathcal{I}}(p_1,p_2,
p_3)=\nonumber\\
             &\sum_\rho \sum_i^{d_F}S^{(\lambda_1)}_{\alpha\alpha'}(p_1) 
S^{(\lambda_2)}_{\beta\beta'}(p_2)
\Psi^\rho_{\alpha\beta\gamma\mathcal{I}}(p_1,p_2,p_3)F^{\rho,i}_{abc}
\end{flalign}
where, for example, for the first term in the antisymmetric component of the 
$\Xi^0$ given above, $S^{(\lambda_1)}$ and $S^{(\lambda_2)}$ would be the 
propagators for $u$- and $s$-quarks, respectively.

\begin{table*}[t!]
 \begin{center}
 \small
\renewcommand{\arraystretch}{1.2}
  \begin{tabular}{lcc}\hline
 state 	& S  &  A   \\ 
 \hline\hline
$p$  & $\frac{1}{\sqrt{2}}\left(udu-duu\right)$  & 
$\frac{1}{\sqrt{6}}\left(2uud-udu-duu\right)$ \\
$n$  & $\frac{1}{\sqrt{2}}\left(udd-dud\right)$  & 
$\frac{1}{\sqrt{6}}\left(udd+dud-2ddu\right)$ \\
$\Sigma^+$ & $\frac{1}{\sqrt{2}}\left(usu-suu\right)$ & 
$\frac{1}{\sqrt{6}}\left(2uus-usu-suu\right)$ \\
$\Sigma^0$ & $\frac{1}{2}\left(usd+dsu-sud-sdu\right)$ & 
$\frac{1}{\sqrt{12}}\left(2uds+2dus-usd-dsu-sud-sdu\right)$ \\
$\Sigma^-$ & $\frac{1}{\sqrt{2}}\left(dsd-sdd\right)$ & 
$\frac{1}{\sqrt{6}}\left(2dds-dsd-sdd\right)$ \\
$\Xi^0$ & $\frac{1}{\sqrt{2}}\left(uss-sus\right)$ & 
$\frac{1}{\sqrt{6}}\left(uss+sus-2ssu\right)$ \\
$\Xi^-$ & $\frac{1}{\sqrt{2}}\left(dss-sds\right)$ & 
$\frac{1}{\sqrt{6}}\left(dss+sds-2ssd\right)$ \\ 
$\Lambda^0$ & $\frac{1}{\sqrt{12}}\left(2uds-2dus+sdu-dsu+usd-sud\right)$ & 
$\frac{1}{2}\left(usd+sud-dsu-sdu\right)$ \\
\hline
\end{tabular}
\caption{Baryon octet flavor amplitudes.
\label{tab:octet_flavor}}
 \end{center}
\end{table*}
\begin{table*}[t!]
 \begin{center}
 \small
\renewcommand{\arraystretch}{1.2}
  \begin{tabular}{lc}\hline
 state 	& S    \\ 
 \hline\hline
$\Delta^{++}$  & $uuu$ \\
$\Delta^{+}$  & $\frac{1}{\sqrt{3}}\left(uud+udu+duu\right)$ \\
$\Delta^{0}$ & $\frac{1}{\sqrt{3}}\left(udd+dud+ddu\right)$ \\
$\Delta^{-}$ & $ddd$ \\
$\Sigma^{*+}$ & $\frac{1}{\sqrt{3}}\left(uus+usu+suu\right)$ \\
$\Sigma^{*0}$ & $\frac{1}{\sqrt{6}}\left(uds+usd+dus+dsu+sud+sdu\right)$ \\
$\Sigma^{*-}$ & $\frac{1}{\sqrt{3}}\left(dds+dsd+sdd\right)$ \\ 
$\Xi^{*0}$ & $\frac{1}{\sqrt{3}}\left(uss+sus+ssu\right)$ \\
$\Xi^{*-}$ & $\frac{1}{\sqrt{3}}\left(dss+sds+ssd\right)$ \\
$\Omega^{-}$ & $sss$ \\
\hline
\end{tabular}
\caption{Baryon decuplet flavor amplitudes.
\label{tab:decuplet_flavor}}
 \end{center}
\end{table*}

We wish to make use of (\ref{eq:spin_flavor_transformation}) to relate the first 
two terms in (\ref{eq:symbolic_faddeev_original}) to the third one, which has 
simpler kinematics and is hence easier to calculate. First, it is necessary to 
realize that if the third term in the equation is evaluated for relative momenta 
$\{p,q\}$ then, considering the permutation of momenta in 
(\ref{eq:spin_flavor_transformation}), the kinematics of the first term is the 
same as that of the third term but evaluated at the transformed momenta 
$\{p',q'\}$
\begin{widetext}
\begin{flalign}
       (p_1=-q-\frac{p}{2}+\frac{P}{3})&\equiv(p'_3=p'+\frac{P}{3}) \nonumber\\
       (k_2=q-\frac{p}{2}+\frac{P}{3}-k)&\equiv 
(k'_1=-q'-\frac{p'}{2}+\frac{P}{3}-k)\quad\Rightarrow\quad 
p'=-q-\frac{p}{2}~,\quad q'=-\frac{q}{2}+\frac{3p}{4}~,\nonumber \\
       (\widetilde{k}_3=p+\frac{P}{3}+k)&\equiv 
(\widetilde{k}'_2=q'-\frac{p'}{2}+\frac{P}{3}+k)
       \label{eq:permutation_delta2}
\end{flalign}
\end{widetext}
and the kinematics of the second term is the same as that of the third term but 
evaluated at the transformed momenta $\{p'',q''\}$
\begin{widetext}
\begin{flalign}
(\widetilde{k}_1=-q-\frac{p}{2}+\frac{P}{3}+k)&\equiv(\widetilde{k}
''_2=q''-\frac{p''}{2}+\frac{P}{3}+k) \nonumber\\
       (p_2=q-\frac{p}{2}+\frac{P}{3})&\equiv 
(p''_3=p''+\frac{P}{3})\hspace*{2cm}\Rightarrow\quad p''=q-\frac{p}{2}~,\quad 
q''=-\frac{q}{2}-\frac{3p}{4}~.\nonumber \\
       (k_3=p+\frac{P}{3}-k)&\equiv 
(k''_1=-q''-\frac{p''}{2}+\frac{P}{3}-k)\label{eq:permutation_delta3}
\end{flalign}
\end{widetext}
Putting in all the elements defined above and permuting the indices in 
(\ref{eq:symbolic_faddeev_original}) as in 
(\ref{eq:spin_flavor_transformation}), and after renaming dummy indices 
conveniently, Eq. (\ref{eq:symbolic_faddeev_original}) now becomes
\begin{widetext}
  \begin{flalign}\label{eq:faddeev_transformed}   
\Psi^\rho_{\alpha\beta\gamma\mathcal{I}}(p,q,P)={}&C~\mathcal{F}_1^{\rho\rho',
\lambda}\int_k~\Bigl[~K_{\beta\beta',\gamma\gamma'}(k'_1,\widetilde{k}'
_2;k)~\delta_{\alpha\alpha''}~S^{(\lambda_2)}_{\beta'\beta''}(k'_1)~S^{
(\lambda_3)}_{\gamma'\gamma''}(\widetilde{k}'_2)\Psi^{\rho'}_{
\beta''\gamma''\alpha'' \mathcal{I}}(p'^{(3)},q'^{(3)},P)~\Bigr]~+ \nonumber \\
&C~\mathcal{F}_2^{\rho\rho',\lambda}\int_k~\Bigl[~K_{\alpha\alpha',\gamma\gamma'
}(k''_1,\widetilde{k}''
_2;k)~\delta_{\beta\beta''}~S^{(\lambda_1)}_{\gamma'\gamma''}(\widetilde{k}
''_1)~S^{(\lambda_3)}_{\alpha'\alpha''}(k''_2)\Psi^{\rho'}_{
\gamma''\alpha''\beta'' \mathcal{I}}(p''^{(3)},q''^{(3)},P)~\Bigr]~+\nonumber\\
&C~\mathcal{F}_3^{\rho\rho',\lambda}\int_k~\Bigl[~K_{\alpha\alpha',\beta\beta'}
(k_1,\widetilde{k}
_2;k)~\delta_{\gamma\gamma''}~S^{(\lambda_1)}_{\alpha'\alpha''}(k_1)~S^{
(\lambda_2)}_{\beta'\beta''}(\widetilde{k}_2)\Psi^{\rho'}_{
\alpha''\beta''\gamma''\mathcal{I}}(p^{(3)},q^{(3)},P)~\Bigr]~,                  
  \end{flalign}
\end{widetext}
where the contraction of the color parts of the kernel and those of the Faddeev 
amplitudes (before the permutation of indices) gives a global factor $C$ and, 
similarly, the contraction of the corresponding flavor parts leads to the flavor 
matrices 
\begin{flalign}\label{eq:flavor_matrices}
\mathcal{F}_1^{\rho\rho',\lambda}=&F^{\dag~\rho}_{bac}k^F_{bb'cc'}F^{\rho',
\lambda}_{b'c'a}~,\nonumber\\
\mathcal{F}_2^{\rho\rho',\lambda}=&F^{\dag~\rho}_{bac}k^F_{aa'cc'}F^{\rho',
\lambda}_{c'a'b}~,\\
\mathcal{F}_3^{\rho\rho',\lambda}=&F^{\dag~\rho}_{bac}k^F_{aa'bb'}F^{\rho',
\lambda}_{a'b'c}~.\nonumber
\end{flalign}
If we denote the result of the integral in the third line of last equation as 
$\left[\Psi^{(3)}_{\lambda_1\lambda_2}\right]^{\rho}_{\alpha\beta\gamma 
\mathcal{I}}(p,q,P)$, it is clear that if the kernel is such that 
$K_{\alpha\alpha',\beta\beta'}(k_1,k_2,k)=K_{\beta\beta',\alpha\alpha'}(k_1,k_2,
k)$ then we have
\begin{flalign}\label{eq:faddeev_transformed_compact}
   \Psi^\rho_{\alpha\beta\gamma 
\mathcal{I}}(p,q,P)={}&\nonumber\\       
C~\mathcal{F}_1^{\rho\rho',\lambda}&\left[\Psi^{(3)}_{\lambda_2\lambda_3}\right]
^{\rho'}_{\beta\gamma\alpha 
\mathcal{I}}(p',q',P)~+\nonumber\\C~\mathcal{F}_2^{\rho\rho',\lambda}&\left[
\Psi^{(3)}_{\lambda_1\lambda_3}\right]^{\rho'}_{\gamma\alpha\beta 
\mathcal{I}}(p'',q'',P)~+\nonumber\\C~\mathcal{F}_3^{\rho\rho',\lambda}&\left[
\Psi^{(3)}_{\lambda_1\lambda_2}\right]^{\rho'}_{\alpha\beta\gamma\mathcal{I}}(p,
q,P)~.
\end{flalign}
Therefore, the problem has beeen reduced to the calculation of only one of the 
diagrams in the Faddeev equation for all possible combinations of pairs of 
quarks. If we rewrite (\ref{eq:faddeev_transformed_compact}) in terms of the 
coefficients $f$ of the expansion of the Fadddeev amplitudes in a covariant 
basis (\ref{eq:basis_expansion}) we finally obtain Eq. (\ref{eq:Faddeev_coeff}) 
with 
\begin{flalign}\label{eq:rotation_matrices} 
H_1^{ij}&=\left[\bar{\tau}^i_{\beta\alpha\mathcal{I}\gamma}(p,q,P)\tau^j_{
\beta\gamma\alpha\mathcal{I}}(p',q',P)\right]~,\\
H_2^{ij}&=\left[\bar{\tau}^i_{\beta\alpha\mathcal{I}\gamma}(p,q,P)\tau^j_{
\gamma\alpha\beta\mathcal{I}}(p'',q'',P)\right]~.
\end{flalign}

\section{Flavor amplitudes}\label{sec:flavor}

For convenience, we reproduce here the usual quark-model flavor amplitudes 
of baryons $F_{abc}$.
The octet is composed of a superposition of states which are symmetric (S) or 
antisymmetric (A) under 
the exchange of the first two indices. In terms of quarks these combinations are 
shown in Tables \ref{tab:octet_flavor} and \ref{tab:decuplet_flavor}.

\end{document}